\documentclass[twocolumn,showpacs,preprintnumbers,amsmath,amssymb,superscriptaddress]{revtex4-1}

\usepackage{graphicx}
\usepackage{dcolumn}
\usepackage{bm}
\usepackage{color}
\usepackage{ulem}
\usepackage{pifont}
\usepackage{natbib}
\usepackage{hyperref}
\hypersetup{
colorlinks = true,
urlcolor   = blue,
linkcolor  = blue,
citecolor  = blue
}

\begin{document}

\preprint{APS/PRB}
\title{Spin dynamics in the Van der Waals magnet CrCl$_3$}

\author{Ola Kenji Forslund}
    \email{okfo@kth.se}
\affiliation{Department of Applied Physics, KTH Royal Institute of Technology, SE-106 91 Stockholm, Sweden}
\author{Konstantinos~Papadopoulos}
\affiliation{Department of Physics, Chalmers University of Technology, SE-41296 G\"oteborg, Sweden}
\author{Elisabetta~Nocerino}
\affiliation{Department of Applied Physics, KTH Royal Institute of Technology, SE-106 91 Stockholm, Sweden}
\author{Gaia~Di~Berardino}
\affiliation{Department of Applied Physics, KTH Royal Institute of Technology, SE-106 91 Stockholm, Sweden}
\author{Chennan~Wang}
\affiliation{Paul Scherrer Institute, Laboratory for Muon Spin Spectroscopy, CH-5232 PSI Villigen, Switzerland}
\author{Jun~Sugiyama}
\affiliation{Neutron Science and Technology Center, 
Comprehensive Research Organization for Science and Society (CROSS), Tokai, Ibaraki 319-1106, Japan}
\affiliation{Advanced Science Research Center, Japan Atomic Energy Agency, Tokai, Ibaraki 319-1195, Japan}
\author{Daniel~Andreica}
\affiliation{Faculty of Physics, Babes-Bolyai University, 400084 Cluj-Napoca, Romania}
\author{Alexander~N.~Vasiliev}
\affiliation{Lomonosov Moscow State University, Moscow, Russia, 119991}
\affiliation{National University of Science and Technology “MISiS”, Moscow, Russia, 119049}
\author{Mahmoud Abdel-Hafiez}
\affiliation{Department of Physics and Astronomy, Uppsala University, Ångströmlaboratoriet, SE-75120 Uppsala, Sweden}
\author{Martin~M\aa nsson}
    \email{condmat@kth.se}
\affiliation{Department of Applied Physics, KTH Royal Institute of Technology, SE-106 91 Stockholm, Sweden}
\author{Yasmine~Sassa}
 \email{yasmine.sassa@chalmers.se}
\affiliation{Department of Physics, Chalmers University of Technology, SE-41296 G\"oteborg, Sweden}

\date{\today}

\begin{abstract}
The magnetic nature of low dimensional compound, CrCl$_3$, was investigated by muon spin rotation, relaxation and resonance ($\mu^+$SR). The $\mu^+$SR measurements revealed three distinct phases as a function of temperature: an antiferromagnetic state  (AF) for $T\leq T_{\rm N}=14.32(6)$~K, a ferromagnetic short range ordered state (FM-SRO) for $T_{\rm N}<T<\sim18$~K and a paramagnetic phase (PM) above $\sim18$~K. Moreover, the AF state exhibits appreciable spin dynamics, which increases with decreasing temperature below $T_{\rm N}$. These dynamics originate from out of plane fluctuations, which seem to settle for $9.5$~K$\leq T\leq T_{\rm N}$, evidenced from measurements in ZF and complementary local field calculations. Moreover, the presented muon Knight shift measurements just above $T_{\rm N}$ represent a clear microscopic evidence for the absence of the previously speculated long range quasi-2D FM order. 




\end{abstract}


\keywords{short keywords that describes your article}

\maketitle

Low-dimensional systems are model materials in which intertwined electronic degrees of freedom lead to strongly correlated ground states. Experimental realizations includes fabrication of quantum/nano dots \cite{Daniel1998, Li2012}, thin films \cite{Venables2000, Geim2010} and even bulk samples \cite{Kimura2008, Forslund2019}, for which the crystal structure may facilitate low dimensional character. These systems where superconductivity \cite{Lange2003}, metal-insulator transition \cite{Mott1968, Kobayashi2019}, spin liquids \cite{Zhou2017} are only a few of many phenomena reported, which have raised the interest of scientists from both fundamental and applied sciences.

Cr$X_3$ ($X =$ I, Cl) are a series of compounds exhibiting low dimensionality. These materials have a rhombohedral symmetry, consisting of 2D Cr layers arranged in a honeycomb web fashion, surrounded by octahedrally coordinated $X$ ions. Many studies on Cr$X_3$ ($X =$ I, Cl) were performed back when the Beatles or ABBA were still on tour \cite{Hansen1958, Cable1961, Narath1964, Kuhlow1982}, but the materials have regained attention due to current topical interest in 2D layered materials \cite{Balleste2011}. Notably, a recent study has shown that monolayers may be obtained via exfoliation \cite{Huang2017}. The interest is driven by the possibility to study low dimensional magnetism, and prospects of electrical control of magnetism for future functional devices \cite{Matsukura2015, Jiang2018}. 



\begin{figure*}[ht]
  \begin{center}
     \includegraphics[keepaspectratio=true,width=160 mm]{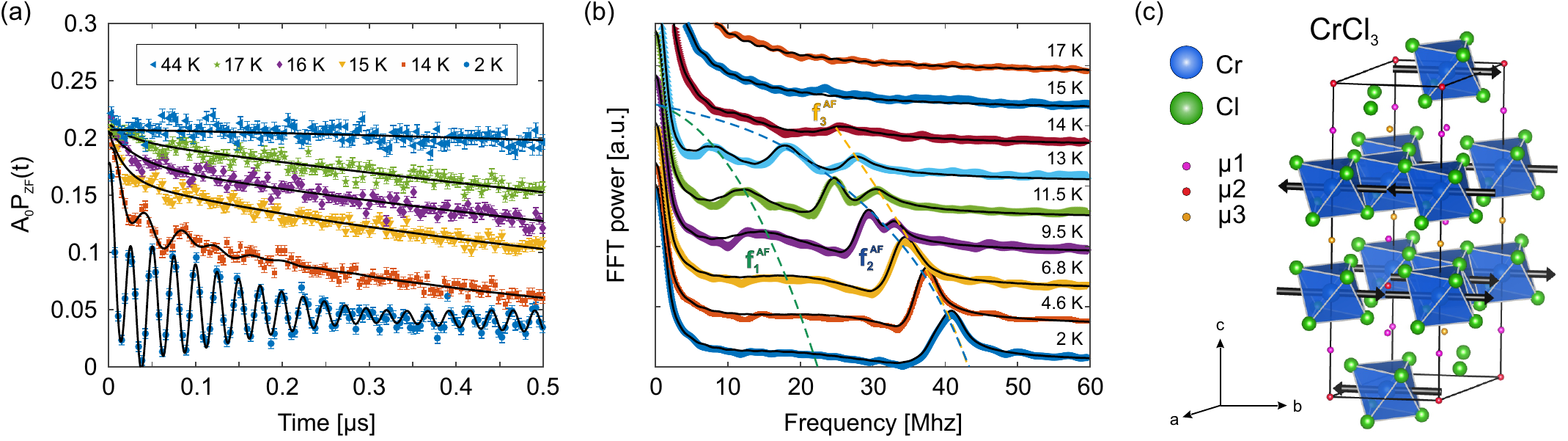}
 \end{center}
    \caption{(a) Zero field (ZF) $\mu^+$SR spectra for selected temperatures ($T = 2$, 14, 15, 16, 17, 19, 24 and 44 K) for the CrCl$_3$ compound. (b) The FFT of the ZF time spectrum as a function of selected temperature. Solid lines represent the best fit using Eq.~(\ref{eq:ZF}). The temperature dependencies of $f^{\rm AF}_1$, $f^{\rm AF}_2$ and $f^{\rm AF}_3$ are highlighted as dashed lines for guide to the eyes. (c) The magnetic structure of CrCl$_3$ drawn within the crystal unit cell, where the ordered moments are shown as black arrows. The predicted muon sites are included as purple ($\mu1$ = (0, 0, 0.15)), red ($\mu2$ = (0, 0, 0)) and orange ($\mu3$ = (0, 0, 0.5)) spheres.}

    \label{fig:ZFSpec}
\end{figure*}

Neutron diffraction measurements at low temperatures revealed that the Cr ions in CrCl$_3$ are ferromagnetically (FM) coupled in the plane and antiferromagnetically (AF) along the c-axis (Fig.~\ref{fig:ZFSpec}(c)) \cite{Cable1961}. A Faraday rotation study \cite{Kuhlow1982} as a function of temperature suggested a rather bold yet fascinating ordering phenomena. On cooling from the paramagnetic state, a first magnetic transition occurs at 16.8~K, at which the Cr spins order ferromagnetically (FM) in a quasi-2-dimensional fashion but they remain disordered in-between the layers. Upon further cooling, the FM structure is rearranged such that an interlayer antiferromagnetic (AF) coupling is stabilised at 15.5~K (AF - quasi-2D FM - PM transition). \cite{Kuhlow1982} This cascade of transitions was also studied by SQUID \cite{Bykovetz2019} and heat capacity \cite{McGuire2017} measurements. The studies support the two transitions scenario but without fully able to experimentally disregard the quasi-2D FM scenario. While the SQUID study hinted at a third possible transition (at 16.8~K, 16.0~K and 14.3~K), such picture was only supported in the low field SQUID measurements suggesting that the magnetism of CrCl$_3$ is sensitive to externally applied magnetic field. This suggests that studies in zero field (ZF) are imperative and relevant for the title compound. 

In this letter, we report the results of a muon spin rotation, relaxation and resonance ($\mu^+$SR) study of CrCl$_3$. The internal magnetic field distribution at the muon site is measured in ZF and compared with the expected one for the reported magnetic structure \cite{Cable1961}. The low temperature out of plane fluctuations are suppressed above above 9.5~K, which seem to alter the magnetic structure. Moreover, transverse field (TF) measurements highlight that a significant local spin density compared with the paramagnetic (PM) state is not established in the intermediate quasi-2D FM state, questioning the previous assessments. Instead of quasi-2D FM, our results point toward the formation of ferromagnetic short range order (FM-SRO), settling the debate. 






Chemical vapor transport (CVT) method was used to prepare single crystals of CrCl$_3$, which were later crushed into a powder form since each individual crystals were not large enough for the experiment. The quality of the sample was checked with basic characterisations prior to the $\mu^+$SR measurements. Details regarding the synthesis is found in SM. The $\mu^+$SR measurements were performed at the surface muon beamline GPS \cite{GPS} at PSI (Switzerland) whereas the DFT calculation were performed using the pseudopotential-based plane-wave method as implemented in $Quantum~Espresso$ \cite{QE-2009, QE-2017}. Details of sample synthesis and experimental setup is found in Ref.~\onlinecite{McGuire2017} and Supplementary materials (SM). 

The collected zero field (ZF) time spectra for selected temperatures, and the corresponding Fourier transform frequency spectra, are shown in Fig.~\ref{fig:ZFSpec}. In order to account for all processes described in the whole measured temperature range, the ZF time spectra were fitted using a combination of oscillations, a stretched exponential, and an exponentially relaxing static Gaussian Kubo-Toyabe (KT) function:

\begin{eqnarray}
 A_0 \, P_{\rm ZF}(t) &=&
\sum_{i}^{3} A^{\rm AF}_i \cos(f^{\rm AF}_i 2\pi t+\phi^{\rm AF}_i)e^{-\lambda^{\rm AF}_i t}\cr &+& A_{\rm tail} e^{-(\lambda_{\rm tail}t)^{\beta_{\rm tail}}} + A_{\rm KT}G^{\rm SGKT}(\Delta_{\rm KT}, t)e^{-\lambda_{\rm KT} t},
\label{eq:ZF}
\end{eqnarray}

where $A_{0}$ is the initial asymmetry determined by the instrument and $P_{\rm ZF}$ is the muon spin polarization function in ZF configuration. $A^{\rm AF}_{i}$, $f^{\rm AF}_{i}$, $\phi^{\rm AF}_{i}$ and $\lambda^{\rm AF}_{i}$ are the asymmetry, frequency, phase and depolarization rate resulting from the internal magnetic field components that are perpendicular with respect to the initial muon spin polarisation. $A_{\rm tail}$, $\lambda_{\rm tail}$ and $\beta_{\rm tail}$ on the other hand are the asymmetry, the relaxation rate and the stretched exponent originating from internal magnetic field components that are parallel with respect to the initial muon polarisation. This form is selected to represent three tail components for $A^{\rm AF}_1$, $A^{\rm AF}_2$, and $A^{\rm AF}_3$ by one term. In a perfect powder, on average, in a perfect powder, on average, 2/3 of the internal fields are oriented perpendicular to the initial muon spin while 1/3 are oriented along the initial muon spin. Indeed, $A_{\rm tail}\simeq \sum A^{\rm AF}_{i}/2$, justifying the usage of a stretched exponential as the tail. $A_{\rm KT}$ and $\lambda_{\rm KT}$ are the asymmetry and the relaxation rate of the KT. The static Gaussian KT is represented by $G^{\rm SGKT}(\Delta_{\rm KT}, t)$ where $\Delta_{\rm KT}$ is related to the second moment of the (Gaussian) field distribution created by the nuclear magnetic moments, considered static in the time window of the experiment. The detailed fitting procedure is outlined and justified in SM.

\begin{figure}[ht]
  \begin{center}
     \includegraphics[keepaspectratio=true,width=85 mm]{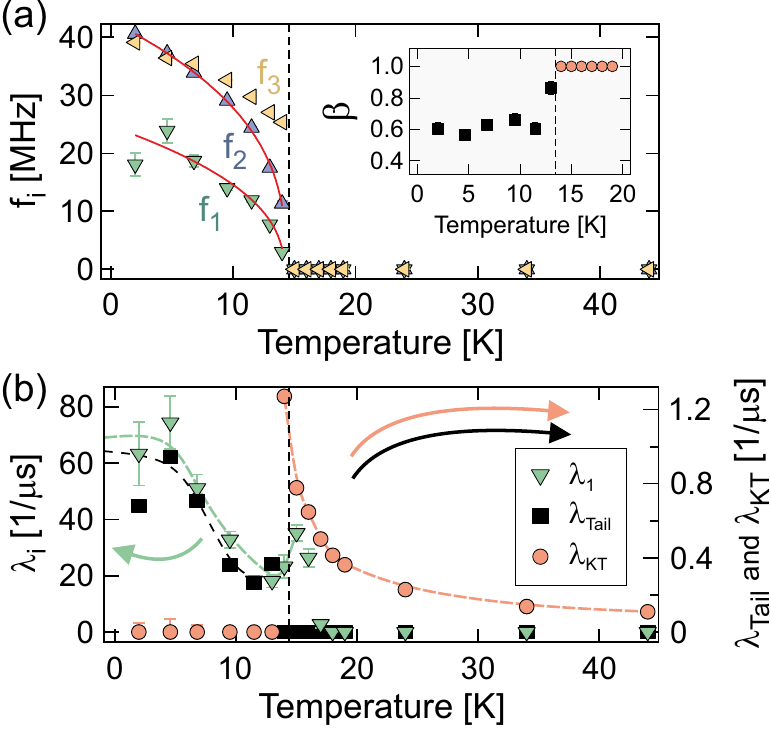}
 \end{center}
    \caption{Temperature dependencies of Zero field (ZF) fit parameters, obtained using Eq.~(\ref{eq:ZF}): (a) precession frequencies ($f^{\rm AF}_1$, $f^{\rm AF}_2$ and $f^{\rm AF}_3$), (b) relaxation rates (right: $\lambda_{\rm tail}$ and $\lambda_{\rm KT}$) and left: $\lambda^{\rm AF}_{1}$. The inset in (a) is the stretched exponent ($\beta_{\rm tail}$). The solid lines in (a) represents the best fit using $f=f_0(1-\frac{T}{T^{\rm ZF}_{\rm N}})^{\alpha}$: $\alpha_1=0.43(14)$, $T^{\rm ZF}_{\rm N}=14.2(4)$ and $f_0=25(3)$ for the signal $f^{\rm AF}_1$ and $\alpha_2=0.36(1)$, $T^{\rm ZF}_{\rm N}=14.32(6)$ and $f_0=42.9(4)$ for the signal $f^{\rm AF}_2$. The dashed lines are guide to the eyes. For clarity, $\lambda^{\rm AF}_2$ and $\lambda^{\rm AF}_3$ are omitted in Fig.~\ref{fig:ZFPara} and presented in SM instead.}
    \label{fig:ZFPara}
\end{figure}

Temperature dependencies of the precession frequencies are shown in Fig.~\ref{fig:ZFPara}(a). Order parameter-like behaviour is observed for $f_{1}$ and $f_{2}$ up to $T^{\rm ZF}_{\rm N}$. In fact, $f_{1}(T)$ and $f_{2}(T)$ can be fitted in accordance with mean field theory: $f=f_0(1-\frac{T}{T^{\rm ZF}_{\rm N}})^{\alpha}$, where $\alpha_1=0.43(14)$, $T^{\rm ZF}_{\rm N}=14.2(4)$~K and $f_0=25(3)$~MHz for the signal $f^{\rm AF}_1$ and $\alpha_2=0.36(1)$, $T^{\rm ZF}_{\rm N}=14.32(6)~$K and $f_0=42.9(4)$~MHz for the signal $f^{\rm AF}_2$. Restricting the fit to closer to $T^{\rm ZF}_{\rm N}$ does not significantly modify the obtained parameters. While both signals are within the error bars of each other, the larger error bars for the $f^{\rm AF}_1$ component is due to the very large value of $\lambda^{\rm AF}_1$, suggesting that the values obtained for $f_2$ are more accurate and we shall define $T_{\rm N}=T^{\rm ZF}_{\rm N}=14.32(6)$. Interestingly, a third frequency component, $f^{\rm AF}_{3}$ presents itself above 9.5~K, which may or may not be present already at 2~K (Fig.~\ref{fig:ZFSpec}(b)). Although, a deviation in the temperature dependence is observed in the $f^{\rm AF}_{3}$ component above 9.5~K and exhibits a rather abrupt change at $T^{\rm ZF}_{\rm N}$. This may suggests that the magnetic structure is altered such that the crystalline moun site is split magnetically above 9.5~K.


The temperature dependencies of the relaxation rates for the slower components ($\lambda_{\rm tail}$ and $\lambda_{\rm KT}$) are shown in Fig.~\ref{fig:ZFPara}(b). $\lambda_{\rm tail}$ corresponds to the spin-lattice relaxation rate and is therefore a measure on how dynamic the system is. A peak like feature is observed between 2 and 9.5~K and is most likely related to difficulty in fitting the $A^{\rm AF}_1$ component, given the high value of $\lambda^{\rm AF}_1$. Instead of a peak feature, one would expect the relaxation rates to increase monotonically with decreasing temperature (see guide to the eyes in Fig.~\ref{fig:ZFPara}(b)). It should be noted that a similar peak behaviour, but less pronounced, is also obtained if two exponentials are used for the tail instead of a streached one. The striking feature is however that the dynamics seems to decrease with increasing temperature, on the contrary to most compounds. Therefore, the internal field dynamics is a driver and an important factor for the system to change its magnetic structure, such that the new magnetic structure splits the crystalline $\mu1$ site into two magnetically different sites, yielding one low frequency and one high frequency precessions. A change in the magnetic structure has not been reported in previous studies \cite{Cable1961, McGuire2017, Bykovetz2019}, suggesting that the change is very subtle. Similarly, a change in the crystal structure has not been reported as well \cite{Cable1961, Morosin1964, McGuire2017}. 

Above $T^{\rm ZF}_{\rm N}$, $\lambda_{\rm KT}$ exhibits a maximum value and rapidly decreases with increasing temperature. The increase at $T^{\rm ZF}_{\rm N}$ corresponds to critical spin fluctuation slowing down, and the decrease with higher temperature suggests an increase in internal magnetic field fluctuations. This temperature dependence is similar to what is found in TF configuration ($\lambda_{\rm TF}$, SM). This behaviour demonstrates spin-spin dynamical correlations in the magnetically non ordered state and is thus expected to follow the temperature dependence of the Curie-Weiss law. Finally, the temperature dependence of the stretched exponent, $\beta_{\rm tail}$, is shown in inset of Fig.~\ref{fig:ZFPara}(a). In this case, $\beta_{\rm tail}\simeq0.6$ at 2~K and increases with increasing temperature. Typically, $\beta$ exhibits values close to 1 at higher temperature and decreases as the temperature is lowered towards 0.3 \cite{Ogielski1985, Campbell1994, Amit1996}. 

Before addressing the quasi-2D FM state, the recorded $\mu^+$SR time spectrum collected at 2~K is reproduced in order to confirm both the proposed magnetic structure and the predicted muon sites. In our past treatment \cite{Forslund2020_Na}, it was shown that the hyperfine contact field is neglectable in an AF, even for an A-type AF. This behaviour is naturally different from that of a FM \cite{Forslund2021_La}. Therefore, the current system shall be modeled assuming only dipolar fields. Based on the magnetic structure of Ref.~\onlinecite{Cable1961}, the internal magnetic field at the considered muon sites can be calculated. Each site yield one single presession frequency. In fact, $\mu1$ and $\mu2$ explain the observed ZF time spectrum collected at 2~K: $\mu1=12.73\simeq17.6(1.7)=f^{\rm AF}_1(2~$K$)$ and $\mu2=39.37\simeq40.64(4)=f^{\rm AF}_2(2~$K$)$. A rather large discrepency is observed for site $\mu1$. The very large field distribution width ($\lambda^{\rm AF}_1$ in Fig.~\ref{fig:ZFPara}(b)) at this temperature is making an exact estimate of the frequency difficult. Expected results are obtained for the $\mu2$ site even without the inclusion of hyperfine contact field. The third component, $f^{\rm AF}_3$, is not reproduced in this calculation. This is explained by; (1) the DFT fails to predict all sites, (2) the determined magnetic structure is not complete or (3) the $f^{\rm AF}_3$ is not present below 9.5~K (Fig.~\ref{fig:ZFSpec}(b)). We further discuss the implications of $f^{\rm AF}_3$ below. 



While $\mu1$ and $\mu2$ explains the main frequencies observed in the data, it does not provide an adequate answer to why the field distribution width of the site $\mu1$ is large ($\lambda^{\rm AF}_1$ in Fig.~\ref{fig:ZFPara}(b)) at low temperatures, and why it becomes structured as the third $f_3$ component becomes significant around 9.5~K. The answer is most likely related to internal magnetic field fluctuations, that seems to be especially active below 9.5~K. Let us first remind ourselves that $\lambda_{\rm tail}$ corresponds to the spin-lattice relaxation rate. This rate corresponds to the rate in which the muon relaxes between the two Zeeman state present for the spin-1/2 particle.
Now, $\lambda_{\rm tail}(2~$K$)\neq0~\mu$s$^{-1}$ was obtained irrespective of fitting function, $i.e.$ a stretched tail or two separate exponential tail, confirming that the ground state is indeed very dynamical. 

Suppose now that the local moments are fluctuating within the a/b plane, given the crystal symmetry and the positions of the muon sites, such fluctuation does not result in significant change in the local field at the $\mu1$ and $\mu2$ sites. However, if the fluctuation are out of plane, the local field at the moun sites changes more dramatically for the $\mu1$ site while this change is not significant for the $\mu2$ site. Therefore, an out of plane fluctuation would explain why $\lambda^{\rm AF}_1$ is large but not $\lambda^{\rm AF}_2$. It seem as these out of plane fluctuations are suppressed as $T_{\rm N}$ is approached and lead to a more static new magnetic structure above 9.5~K. 

The detailed magnetic structure can be deduced from the presented data and model. There are many possible magnetic structure and we have selected few relevant ones. First of, canting the in-plane moments out of plane may result in a more defined internal field at $\mu_1$ site with internal field values closer to experimentally obtained ones. Although, a simple canting does not split the sites magnetically. Similarly, an A-type AF with moments aligned parallel to c-axis provide somewhat reasonable results but it does not split the sites either. In fact, many simple structure results in none split $\mu1$ site. Therefore, if $f^{\rm AF}_3$ is a result of a split of the known crystalline moun sites ($\mu1$ and $\mu2$), the magnetic structure above 9.5~K is most likely a little more complicated. If on the other hand $f^{\rm AF}_3$ is from a separate crystalline muon site not predicted with DFT, we may conclude that the reported magnetic structure \cite{Cable1961} is consistent with the presented data but with a potential canting present below 9.5~K. In this case, the degree of the canting is likely to decrease with increasing temperature as the out of plane fluctuations are suppressed. Either case, the detailed magnetic structure is ideally re-investigated in a detailed neutron diffraction study.

While the system is generally considered as a 2D system, mean field fits of the order parameters (Fig.~\ref{fig:ZFPara}(a)) yielded exponents $>0.36$, suggesting 3D like fluctuations below $T_{\rm N}$ \cite{Blundell2003, Taroni2008}. In fact, neutron diffraction \cite{Cable1961}, spin wave analysis \cite{Narath1965}, heat capacity \cite{McGuire2017} and magnetisation \cite{Bykovetz2019} measurements suggest very weak anisotropy. Relatively weak field is able to spin polarise the low temperature AF phase. Moreover, the obtained effective moment from magnetisation measurements is consistent with a value expected for spin only Cr$^{3+}$. In other words, the orbital contribution is suppressed and anisotropy is thus expected to be small. This assessment is also supported by the TF measurements presented below, as small values of TF easily polarises the system.

Now, lets turn our focus onto the proposed quasi-2D FM phase. Faraday rotation measurements \cite{Kuhlow1982} showed that the sample undergoes two transitions, one at 16.8~K and one around 15.5~K: from a high temperature paramagnetic (PM) phase, to a quasi-2D ferromagnetic (FM) phase that is disordered in three dimensions to finally a 3D antiferromagnetic (AF) phase. While studies so far has not manage to neither confirm or reject the quasi-2D FM scenario, we shall now present microscopic evidence for the absence of a quasi-2D FM order.

\begin{figure}[ht]
  \begin{center}
     \includegraphics[keepaspectratio=true,width=65 mm]{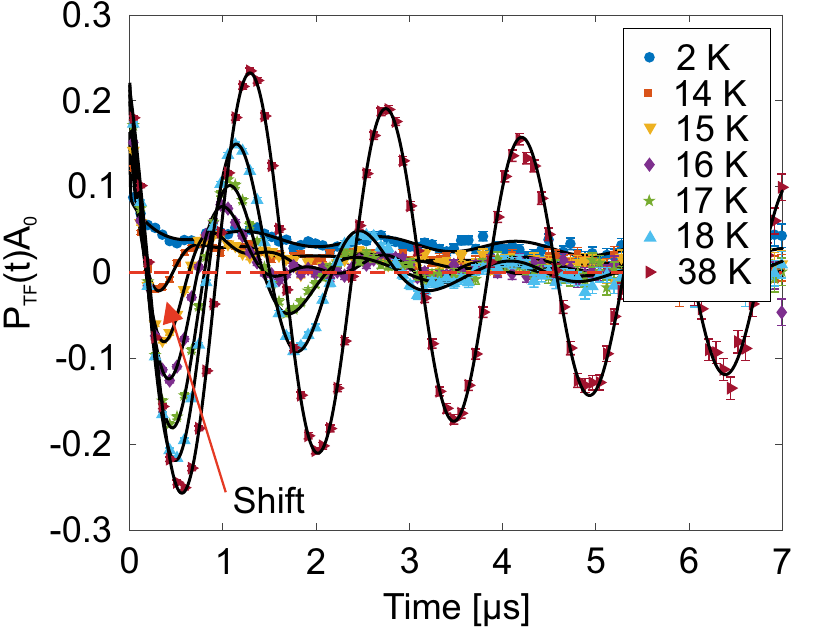}
 \end{center}
    \caption{(a) Transverse field (TF) $\mu^+$SR spectra for selected temperatures ($T=2$, 14, 15, 16, 17, 18 and 38~K) for the CrCl$_{3}$ compound. Solid lines represent the best fit using Eq.~(\ref{eq:TF}).
  }
    \label{fig:TFSpec}
\end{figure}

A clear microscopic evidence of this quasi-2D order can be obtained by measuring the local spin density at the muon site. Therefore, TF as a function of temperature was measured and Fig.~\ref{fig:TFSpec} shows the collected TF $(= 50$~G $\simeq0.6871$~MHz) time spectra for selected temperatures. Two oscillations are needed in order to fit the collected data, due to the presence of the two muon sites ($\mu1$ and $\mu2$). As the temperature is lowered, the frequency is shifted towards higher values, accompanied by a suppression in the amplitude of the oscillation. Therefore, the TF contributions to the time spectra were fitted with three exponentially relaxing oscillations

\begin{eqnarray}
 A_0 \, P_{\rm TF}(t) &=&
\sum_{i}^{2} A^{\rm TF}_{i} \cos(f^{\rm TF}_{i}2\pi t+\phi^{\rm TF}_{i})e^{-\lambda^{\rm TF}_{i}t}\cr &+&A^{\rm TF}_{\rm imp} \cos(f^{\rm TF}_{\rm imp}2\pi t+\phi^{\rm TF}_{\rm imp})e^{-\lambda^{\rm TF}_{\rm imp}t},
\label{eq:TF}
\end{eqnarray}
where $A_{0}$ is the initial asymmetry determined by the instrument and $P_{\rm TF}$ is the muon spin polarization function in TF configuration. $A^{\rm TF}$, $f^{\rm TF}$, $\phi^{\rm TF}$ and $\lambda^{\rm TF}$ are the asymmetry, frequency, phase and depolarization rate resulting from the applied TF, where the superscripts $i$ and imp represent the contributions from the sample and impurity phase ($<3\%$). The asymmetry of the impurity contribution was fixed throughout the whole measured temperature range, a value estimated at 2~K. The internal magnetic field contribution present below $T^{\rm TF}_{\rm N}$ were fitted with a combination of exponentials and is further explained in SM, together with the the detailed fitting procedure. 

Temperature dependence of muon Knight shift for each site can be obtained from fitting the data using Eq.~\ref{eq:TF}, and is defined as

\begin{eqnarray}
K_{\mu,i}=\frac{\bm B_{\rm ext} \cdot (\bm B_{\rm loc}- \bm B_{\rm ext})}{B_{\rm ext}^2}=\frac{f^{\rm TF}_i-f_{\rm ref}}{f_{\rm ref}}
\label{eq:Knight}
\end{eqnarray}

where the value of the precession frequency obtained at highest measured temperature, $f_{\rm ref}=0.6871$~MHz, was used. Naturally, this results in $K=0$ at the highest measured temperature. The applied field yields positive muonic knight shift as $T^{\rm TF}_{\rm N}$ is approached, despite a TF of 50~G. This suggest that the compound is indeed quite susceptible to externally applied magnetic field, and is consistent with previously reported magnetic field dependent studies \cite{Cable1961, McGuire2017, Bykovetz2019}. The positive shift suggest that the local field aligns with the applied field, and a paramagnetic like state is thus expected. 

\begin{figure}[ht]
  \begin{center}
     \includegraphics[keepaspectratio=true,width=80 mm]{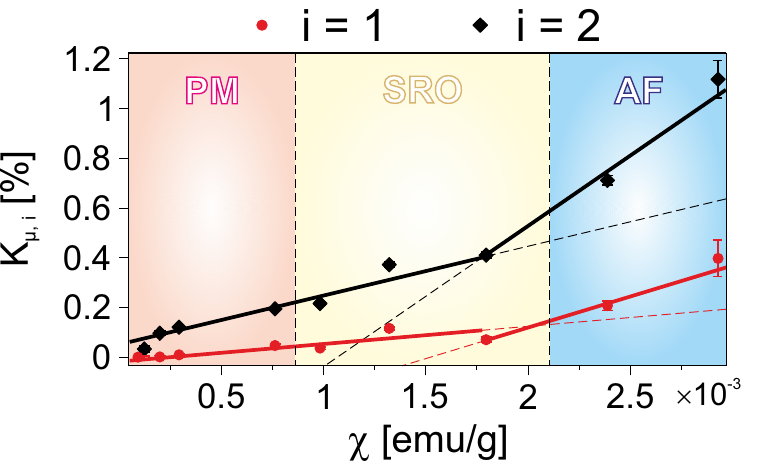}
 \end{center}
    \caption{Muon Knight shift ($K_{\mu,i}$) and the bulk magnetic susceptibility ($\chi$) is plotted with temperature as an implicit parameter. The solid lines represents best linear fits, where the dotted parts are extrapolated. The vertical dashed lines indicted the phase temperature boundaries of the sample: antiferromagnetic (AF), short range order (SRO) and paramagnetic (PM) phases. The AF-SRO transition temperature boundary is around $T=14$~K, while SRO-PM transition temperature is around 17-18~K. The $\chi$ was extracted from a DC-magnetisation measurement under zero field cooling protocol, with an applied of $B=1000$~G.
    }
    \label{fig:KnightX}
\end{figure}

The microscopic origin to the local field, $\bm B_{\rm loc}$, consists of dipolar field, hyperfine contact field and a temperature independent component. Therefore, the temperature dependence can be attributed to coupling between the muon and polarised localised and intenerant electrons. Given that the dipolar and hyperfine contact field contribution is related to the local magnetic susceptibility, the Knight shift is expected to follow a linear behaviour when plotted against the bulk magnetic susceptibility with temperature as an intrinsic parameter, as shown in Fig.~\ref{fig:KnightX}, where linear behaviours are observed for both sites. The linearity is present from higher temperature (PM phase) down to $T^{\rm ZF}_{\rm N}$, the AF-SRO boundary. At this point however, the derivative changes and is most likely attributed to an increase in the hyperfine coupling constant, as the sample forms an AF order. Of course, the increase can be attributed due to the fact that fitting weak-TF becomes difficult close to $T^{\rm TF}_{\rm N}$ because the TF asymmetry decreases (SM). More importantly though, a change in shape is not present between the PM-SRO phases (around $17-18$~K). This suggests that the local field does not ordered in such a way that the local spin density modifies. For a fully ordered quasi-2D FM, one would expect a significant change in the hyperfine coupling, even if it is just ordered in 2D. These results represent clear microscopic evidence for the absence of "quasi-2D FM" in this compound, as proposed by Ref.~\cite{Kuhlow1982}.  Instead of "quasi-2D FM", it is more natural to assign this phase to SRO. While our data cannot determine the detailed interaction, previous studies \cite{Kuhlow1982, McGuire2017} have suggested this phase to be FM like in nature and we may thus assign it to FM-SRO. 



Recently, topological spin excitations were observed in the related compound, CrI$_3$ \cite{Chen2018}, given the 2D honeycomb structure of the Cr. A similar behavior may be expected in CrCl$_3$ as well, provided the similar arrangement of 2D honeycomb Cr layers, surrounded by octahedrally coordinated $X$ ion. Compounds with a nontrivial topological magnon edge states are highly attractive for future dissipationless and highly efficient spintronic applications. Naturally, unveiling the details as presented is imperative before undertaking a detailed study of bulk and/or thin films. 

In summary, the two dimensional Van der waals magnet, CrCl$_3$, was investigated by muon spin rotation, relaxation and resonance ($\mu^+$SR). This study clarified the magnetic phases present in the compound as a function of temperature: an antiferromagnetic state  (AF) for $T\leq T_{\rm N}=14.32(6)$~K, a ferromagnetic short range ordered state (FM - SRO) for $T_{\rm N}<T<\sim18$~K and a paramagnetic phase (PM) for $T\geq\sim18$~K. Zero field (ZF) measurements, complemented by local field calculations, confirmed the magnetic order proposed by neutron diffraction. However, the ground state is found to be highly dynamic, which decreases with increasing temperature, evidenced by the temperature dependence of the spin-lattice relaxation rate. ZF data suggested that the local magnetic structure is changed above 9.5~K, driven by out of plane fluctuations. Knight shifts measurements on the other hand provided with microscopic evidence for the absence of a quasi-2D FM order, which was previously proposed by Faraday measurements. Instead, a ferromagnetic short range order (FM - SRO) is stabilised just above $T_{\rm N}$, for which a paramagnetic phase is recovered above $\sim18$~K.

\begin{acknowledgments} 
We thank the staff of PSI for help with the $\mu^+$SR experiments. This research was supported by the Swedish Research Council (VR) (Dnr. 2016-06955) as well as the Swedish Foundation for Strategic Research (SSF) within the Swedish national graduate school in neutron scattering (SwedNess). Y.S. is funded by the Swedish Research Council (VR) through a Starting Grant (Dnr. 2017-05078). Y.S. and K.P. acknowledge funding a funding from the Area of Advance- Material Sciences from Chalmers University of Technology. D.A. acknowledges partial financial support from the Romanian UEFISCDI Project No. PN-III-P4-ID-PCCF-2016-0112. M.A.H. acknowledges financial support from the Swedish Research Council (VR) under project No. 2018-05393. Support by the P220 program of Government of Russia through the project 075-15-2021-604 is acknowledged. J.S. acknowledge support from the Ministry of Education, Culture, Sports, Science and Technology (MEXT) of Japan, KAKENHI Grant No.23108003 and Japan Society for the Promotion Science (JSPS) KAKENHI Grant No. P18H01863 and JP20K21149. The data was analysed with software package musrfit \cite{musrfit}. The crystal figure was drawn using VESTA \cite{Momma2008}.

\end{acknowledgments}

\bibliography{Refs} 
\end{document}